\documentstyle[12pt]{article}
\newcommand{\chem}[2]{$\rm{}^{#1}\kern-0.8pt#2$}
\newcommand{\chim}[2]{\rm{}^{#1}\kern-0.8pt#2}
\newcommand{\reac}[6]{$\rm\,{}^{#1}\kern-0.8pt{#2}\,({#3}\,,{#4})\,
           {}^{#5}\kern-0.8pt{#6}\,$}
\newcommand{\gsimeq}{\,\,\raise0.14em\hbox{$>$}\kern-0.76em\lower0.28em\hbox  
{$\sim$}\,\,}
\newcommand{\lsimeq}{\,\,\raise0.14em\hbox{$<$}\kern-0.76em\lower0.28em\hbox  
{$\sim$}\,\,}

\newcommand{\ms}{$\rm M_{\odot}$}

\input epsf
\title{\bf WOLF--RAYET STARS AND RADIOISOTOPE PRODUCTION}
\author{G.~Meynet$^{1}$ and M. Arnould$^{2}$\\
\vspace{1cm}\\
\normalsize $^{1}$ Geneva Observatory, CH-1290 Sauverny \\
\normalsize $^{2}$ Institut d'Astronomie et d'Astrophysique, Universit\'e 
            Libre de Bruxelles \\
\normalsize            C.P. 226, Bd. du Triomphe, B-1050 Brussels, Belgium}
\date{\mbox{}}
\begin{document}
\maketitle
\pagestyle{empty}
%
% WE REDEFINE THE plain LaTeX PAGESTYLE !!! 
% THIS PAGESTYLE WILL BE USED FOR THE FIRST PAGE ONLY !
%
\def\bull{\vrule height .9ex width .8ex depth -.1ex}
\makeatletter
\def\ps@plain{\let\@mkboth\gobbletwo
\def\@oddhead{}\def\@oddfoot{\hfil\tiny\bull\quad
``Astronomy with radioactivities'';
Tegernsee, 1999\quad\bull}%
\def\@evenhead{}\let\@evenfoot\@oddfoot}
\makeatother
%
% AND DEFINE OUR MACROS FOR THE REFERENCE LIST
% I.E \beginrefer \refer and \endrefer
%
\def\beginrefer{\section*{References}%
\begin{quotation}\mbox{}\par}
\def\refer#1\par{{\setlength{\parindent}{-\leftmargin}\indent#1\par}}
\def\endrefer{\end{quotation}}
%
% BEGIN THE ABSTRACT CHAPTER WITH \noindent\small, ENCLOSE IT IN A GROUP
% AND BOLDFACE THE TITLE.
%
{\noindent\small{ } 
}
%
% NOW COMES THE MAIN BODY OF THE ARTICLE
%

Radioisotopes are natural clocks
which can be used to
estimate the age of the solar system and of the Universe
(see e.g. Takahashi 1998; Chen \& Tilton 1976; see also the recent review by
Arnould \& Takahashi 1999).
They are responsible for the steady decline of the light
curves of supernovae (see e.g. Diehl \& Timmes 1998, and references therein).
The diffuse emission at 1.8 MeV from the decay of $^{26}$Al (e.g. Arnould \&
Prantzos for a recent review) may also provide a measure of the  present day
nucleosynthetic activity in our Galaxy.  Therefore, even if radionuclides
represent only a tiny fraction of the  cosmic matter, they carry unique pieces 
of information.

A great number of radioisotopes are produced by massive stars
at the  time of the supernova explosion. A fraction of them are also
produced during the previous hydrostatic 
burning phases. These nuclides are then ejected either at the time of the
supernova event, or through stellar winds during the hydrostatic burning
phases.
This paper focusses  on the non-explosive ejection of radionuclides by  
Wolf-Rayet (W--R) stars.

\section{THE W--R STARS}

Wolf-Rayet stars are generally believed to be bare cores of initially massive
stars 
(Lamers et al  1991) whose original H--rich envelope has been removed 
by stellar winds or through a Roche lobe overflow in a close binary system
(see the recent reviews by van der Hucht 1992; 
Maeder \& Conti 1994; Willis 1999).
Observationally, most W--R stars appear to originate from stars initially
more massive than about 40 M$_\odot$ (Conti et al 1983; Conti 1984). Some of
them, however, may have progenitors with initial
masses as low as 15--25 M$_\odot$ (see e.g. Hamann \& Koesterke 1998; 
Massey \& Johnson 1998). 
The stars enter the W--R phase as WN stars whose surface
abundances are representative of equilibrium CNO
processed material. If the peeling off
proceeds deep enough, the star may enter the WC phase,
during which the He--burning products appear at the surface.

Many observed features are well reproduced by current stellar models.
Typically,  good agreement is obtained between the
observed and predicted values for the surface abundances of WN stars
(Crowther et al  1995; Hamann \& Koesterke 1998a).
This  indicates the general 
correctness of our understanding of the CNO cycle (Maeder 1983), but is not a
test of the model structure.
For WC stars,  
comparisons with observed surface abundances also   
show in general a good agreement (Willis 1991; Maeder
\& Meynet 1994). In particular, the strong surface Ne--enrichments 
predicted by the models of WC stars have been confirmed 
by ISO observations (Willis et al 1997, 1998). 

The ratios of the numbers of W--R to O-type stars (W--R/O) or to red
supergiants
(W-R/RSG), as well as the number of WN with respect to WC stars show a
strong correlation with metallicity (Maeder 1991; Maeder \& Meynet 1994;
Azzopardi et al.  1988; Smith 1988; Massey \& Jonhson 1998).
For instance, the W--R/O number ratio increases
with the metallicity $Z$ of the parent galaxy.
The main reason for this trend is the metallicity dependence of the stellar
winds, which have a strong impact on the development of the W--R phase and on
its lifetime (Smith 1973;  Maeder et al.  1980).
The higher the metallicity, the stronger is the mass loss 
by stellar winds and thus the earlier is the entry in
the W--R phase for a given star. In addition,  
the minimum initial mass for forming a W--R star
is lowered. This trend will also be responsible for a greater W--R production
of radioisotopes in high metallicity regions as, for instance, the inner
zones of our Galaxy.

\section{THE $\rm{\gamma}$-RAY LINE CONNECTION}

\chem{26}{Al} is a key radionuclide in the development of  
$\gamma$-ray line astrophysics. This comes from the observation in the
present interstellar medium (ISM) of the 1.8
MeV $\gamma$-ray line emitted following the de-excitation of the \chem{26}{Mg}
produced by the \chem{26}{Al} $\beta$-decay.

The data available to-date indicate that the present ISM contains
about 2 M$_\odot$ of
\chem{26}{Al}, the distribution of which excludes (i) a unique point source in
the
galactic center, (ii) a strong contribution  from the old stellar population of
the
galactic bulge, and (iii) any class of sources involving a large number of
sites with
low individual yields, like novae or low-mass AGB stars. In contrast, they
favor
massive stars (W--R stars and/or SNIb/Ic and SNII) as the
\chem{26}{Al} production sites (e.g. Kn\"odlseder et al. 1999).

 A detailed discussion of the production of \chem{26}{Al} by the MgAl chain of
hydrogen burning developing in non-rotating W--R stars has been conducted
recently by Arnould et al. (1997b)
and Meynet et al. (1997). It appears that the
\chem{26}{Al} yields increase with initial mass and metallicity $Z$, the $Z$
dependence
being
approximated by $M_{\rm 26}(M_{\rm i},Z) = (Z/Z_\odot)^2M_{\rm 26}(M_{\rm
i},Z_\odot)$, where $Z_\odot = 0.02$ is the solar metallicity and
$M_{\rm 26}(M_{\rm i},Z)$ the mass of \chem{26}{Al} ejected by stars
of initial mass $M_{\rm i}$ and metallicity $Z$. It is worth
noticing
that the $^{26}$Al yields of Meynet et al. (1997) are in qualitative agreement
with
those of Langer et al. (1995), even if these two sets of models greatly
differ in their physical ingredients. 

These \chem{26}{Al} yields from non-rotating W--R stars have been used to
evaluate quantitatively the virtues of these
stars as
sources of the 1.8 MeV line in the present ISM. Figure 1 shows the mass of live
$^{26}$Al deposited by the winds of W--R stars in rings of increasing
galactocentric
radius. This estimate is based on the metallicity-dependent yields computed by
Meynet et al. (1997), and on their assumptions concerning in particular the
Initial Mass Function and the galactocentric radius dependence of the
metallicity and star formation rate.
The signature of the
ring of molecular clouds located at the galactocentric radius of
about 5
kpc is clearly seen. It is also predicted that more than half of the total
\chem{26}{Al} mass is contained within this ring. 

The integration of the histogram of Fig. 1 over the galactic radius leads to a
total
galactic mass of 1.15 M$_\odot$ ejected by the stellar winds of non-rotating
W--R
stars. Due consideration of various uncertainties leads to masses in the
probable 
0.4-1.3 M$_\odot$ range (Meynet et al. 1997), so that the considered stars
might account
for 20 to 70\% of the present galactic \chem{26}{Al}. 

\begin{figure}
\setlength{\epsfxsize}{8cm}%
\leavevmode\epsffile{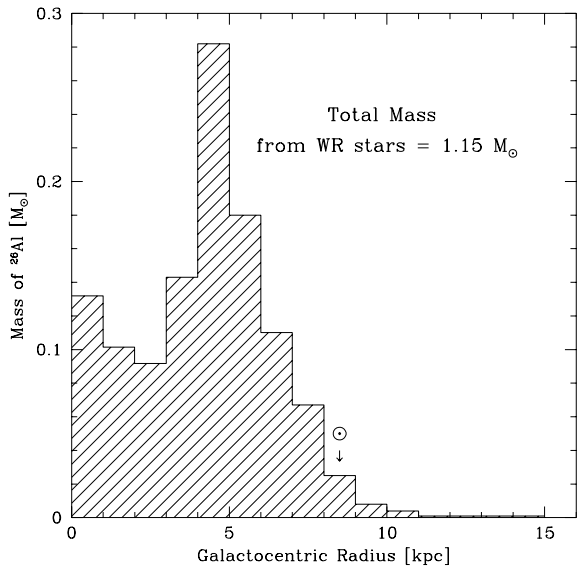}% -- eps file name
\caption{Galactocentric radius dependence of the mass of $^{26}$Al ejected 
by the winds of non-rotating W--R stars. The Galaxy is divided
into 15 concentric 
rings with 1 kpc width.
} \label{fig-3}
\end{figure}

The presently available observations cannot unambiguously disentangle the
relative contributions to
the present \chem{26}{Al} galactic content of the W--R star winds, of the 
W--R SNIb/Ic explosions, and of SNII supernovae. In this respect, a recent
observation of
high value provides an upper limit of about 10$^{-5}$ ph cm$^{-2}$s$^{-1}$ for
the
1.8 MeV luminosity of the $\gamma^2$ Vel binary system containing an
O-type and a W--R star (of WC8 subtype, Oberlack et al. 1999). In order to
evaluate the
compatibility of this observation with predictions, at least the initial mass
and  
metallicity of the W--R progenitor, as well as the age of $\gamma^2$ Vel have
to
be
known. The proximity of 
$\gamma^2$ Vel justifies the use of solar metallicity stellar models.
Non-rotating
single star models (Meynet et al. 1994) combined with the position of the
O-star
component
in the HR diagram lead to an age of about 3.6 10$^6$ yr for the system (De
Marco \& Schmutz 1999), and an initial W--R 
mass of about 60 M$_\odot$. Such a W--R progenitor is
predicted to have a 1.8 MeV luminosity exceeding the observed upper limit by a
factor of about 2. Rotation would lead to a more gradual exposure of
$^{26}$Al at the stellar surface than in the limit of no rotation  (see below).
It remains however to be demonstrated through detailed calculations that
rotation can indeed cure the $\gamma^2$ Vel discrepancy. Let us simply note
that the use of rotating star models to estimate the age of the O-type star
may also increase its age, modifying the possible range of initial masses for
the
W--R companion. The binary nature of 
$\gamma^2$ Vel might  also provide some remedy if indeed a large enough
fraction
of the W--R \chem{26}{Al}-loaded ejecta could be accreted by the O-star
companion.

\begin{figure}
\setlength{\epsfxsize}{8cm}%
\leavevmode\epsffile{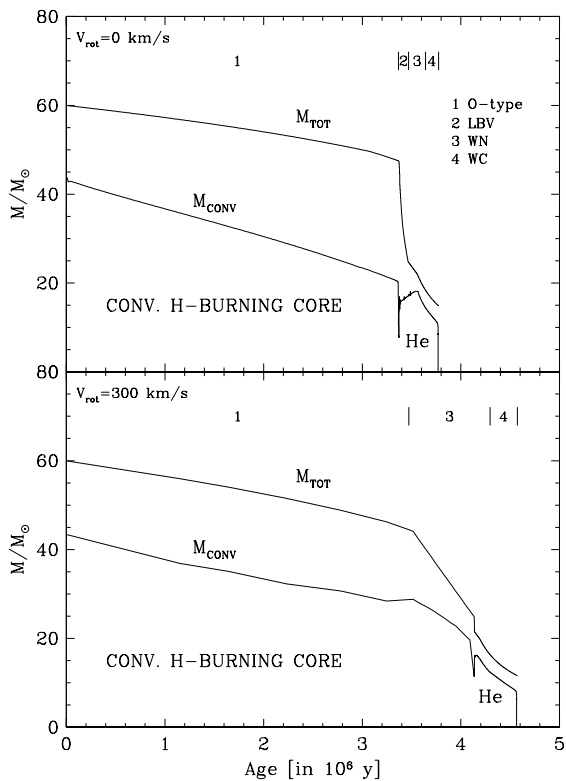}% -- eps file name
\caption{Time evolution of the total mass $M_{\rm TOT}$ and of the mass of the
convective core $M_{\rm CONV}$. Various evolutionary stages are indicated at
the top of the
figure.
} \label{fig-2}
\end{figure}

\subsection{Effect of rotation on the W--R production of\, \chem{\bf 26}{\bf
Al}}

Rotation induces many dynamical instabilities in stellar interiors, and in
particular meridional circulation and shear
turbulence (e.g. Zahn 1992). The related mixing of the chemical species can
deeply modify the chemical structure of a star and its evolution, and may in
particular have important consequences on the $^{26}$Al
production by W--R stars. At present, very few rotating W--R models address
this
question
in detail (see Langer et al. 1995 for some preliminary results), so that  
it is premature to quantitatively assess the possible role played by rotation
in
that respect. However, it seems safe to say that rotation increases the
quantity
of
$^{26}$Al ejected by W--R stellar winds. This claim is made plausible by Fig.
2,
which
shows the structural evolution during their H- and He-burning phases of two 60
M$_\odot$ stars that just differ by their rotational velocities. It appears
that 

\noindent 1) The size of the convective core is increased by rotation;

\noindent 2) During the O-type star phase, the total mass decreases faster in
the rotating model; 

\noindent 3) Rapidly rotating stars may enter the W--R phase while still on the
Main
Sequence. Moreover the surface abundances characteristic
of their WNL and WC phases are not due to the mass loss
which uncovers core layers, but result from diffusive mixing in the
radiative zones. As a consequence, the evolution of the surface abundances are
much smoother in rotating models (see Fig. 3);

\noindent 4) The W--R lifetime increases with the initial rotational velocity.
In
particular the WN stage during which $^{26}$Al is ejected lasts much longer,
so that much more mass is
ejected during this phase. As a numerical example, the non-rotating model
sketched in Fig. 2 ejects about 7 M$_\odot$ during the WN phase, while about 28
M$_\odot$ are ejected by the rotating model during the same phase.
 
The inference that rotation can enhance the \chem{26}{Al} yields of W--R stars
is
confirmed by the numerical simulations of Langer et al. (1995). Rotation also
lowers the
minimum initial mass of single stars which can go through a W--R stage with a
concomitant ejection of \chem{26}{Al}. This likely contributes to the net
amount
of \chem{26}{Al} ejected by W--R stars in the ISM, which may be of relevance
both
to
cosmochemistry and to $\gamma$-ray line astrophysics.

\begin{figure}
\setlength{\epsfxsize}{15cm}%
\leavevmode\epsffile{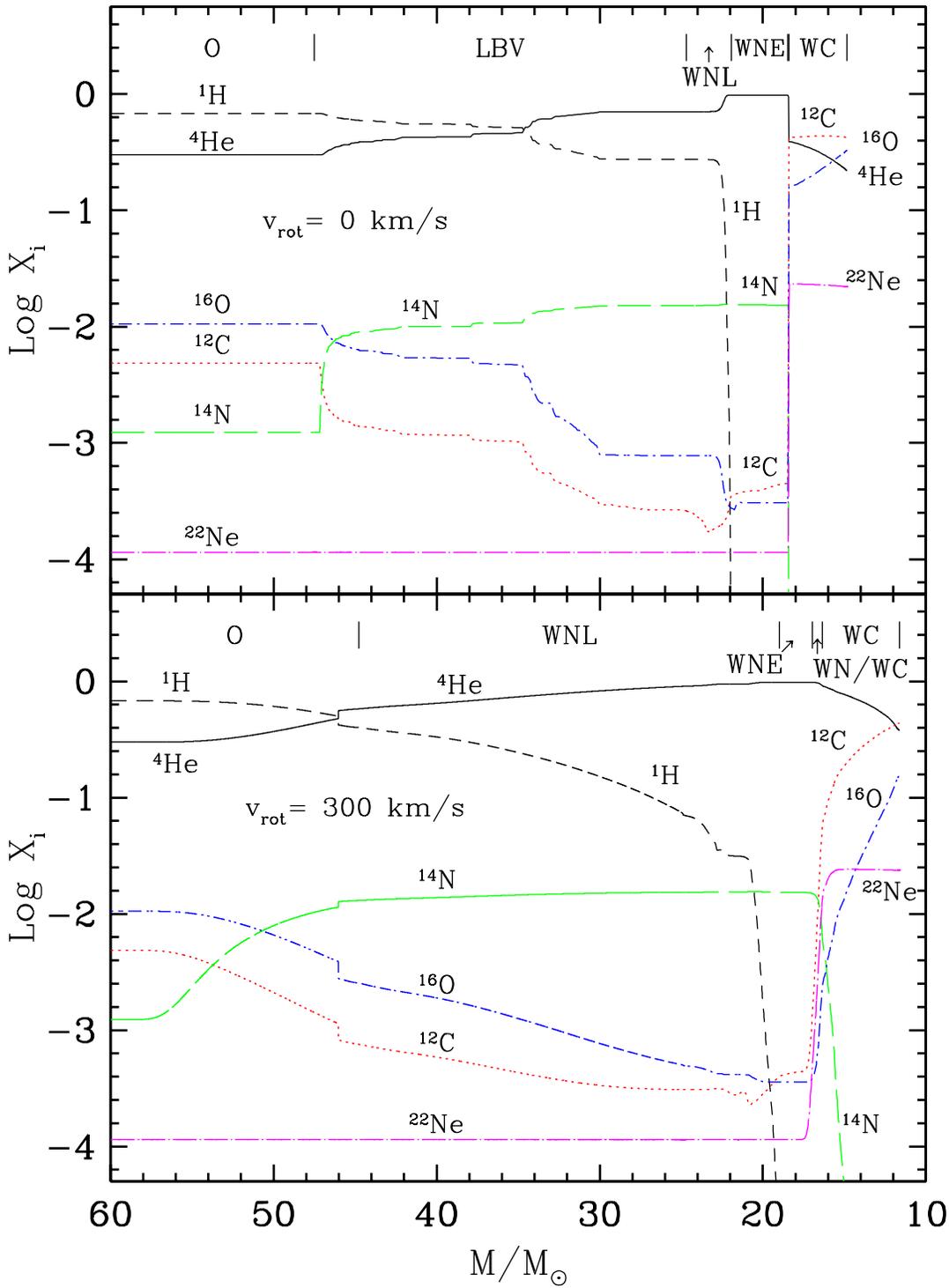}% -- eps file name
\caption{Evolution of the abundances at the surface of a
60 M$_\odot$ star as a function of the remaining stellar mass
for different initial rotational velocities $v_{\rm rot}$. 
The portions
of the evolution during which the star may be considered
as an O-type star, a LBV and a W--R star are indicated. During
the W--R phase, the WN, the transition ``WN/WC'' and 
the WC phases are identified.
} 
\end{figure}

\begin{figure}
\setlength{\epsfxsize}{8cm}%
\leavevmode\epsffile{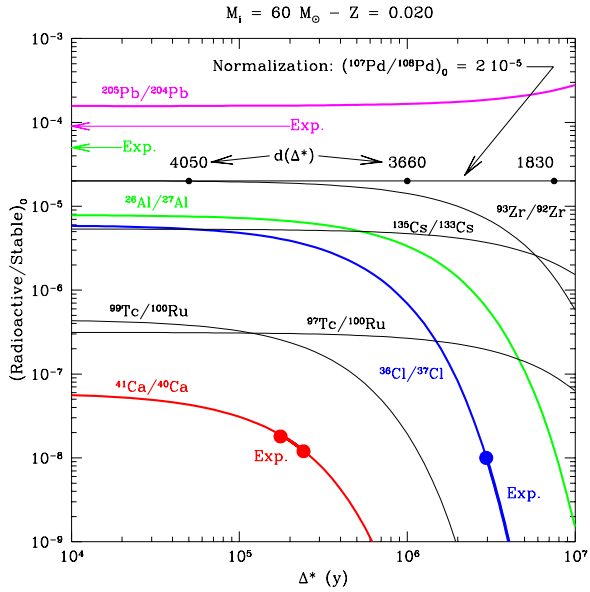}% -- eps file name
\caption{Abundance ratios $\mbox{(R/S)}_0$ of various radionuclides R relative
to
stable neighbours S versus $\Delta^\ast$ (see main text) for the 
60 M$_\odot$ model star with $Z = 0.02$. All the displayed ratios are
normalized to
$(\chim{107}{Pd}/\chim{108}{Pd})_0 = 2\,10^{-5}$  (e.g. Wasserburg 1985)
through
the application of a common dilution factor $d(\Delta^\ast$). The values of
this
factor are indicated on the Pd horizontal line for 3 values of
$\Delta^\ast$. Other available experimental data (labelled Exp) are displayed.
They are adopted from MacPherson et al. (1995) for Al, Srinivasan et al. (1994)
for Ca, Murty et al. (1997) for Cl, and Huey \& Kohman  (1972) for Pb (see
Arnould
et al. 1997ab for more details)
}
\end{figure}

\subsection{Effect of binarity on the W--R production of \chem{\bf 26}{\bf Al}}

Tidal interactions in close binary systems may considerably modify the
evolution of the two stellar components with respect to the one they would
experience as isolated stars. Mass transfer by
Roche Lobe Overflow can mimic a strong stellar wind, and thus reduce the
critical
initial mass for producing single W--R stars. Before any mass transfer, tidal
effects
are also expected to deform the star and therefore induce instabilities
reminiscent of those induced by rotation. To our knowledge, the latter effect
has
never been studied in any detail, even if it might have important consequences,
like
the homogenization of the stars, and the related inhibition of 
mass transfer. Other effects remain to be explored, like the impact of
colliding
winds on the mass transfer process, or even the very nature of the grains which
can condense in colliding winds. Thus the effect
of binarity on the W--R star formation and evolution, and on the corresponding
$^{26}$Al production cannot be evaluated with confidence at this time. Some
preliminary estimates (Braun \& Langer 1995) lead to the conclusion that `only
for
stars
with masses
$\le$ 40 M$_\odot$, binarity has the potential to increase the $^{26}$Al yield
compared to the single star case'. The fact that the
situation may be quite different in more massive stars can be interpreted in
the
following way: after their Main Sequence, single high mass stars go through a
LBV
phase characterized by especially high stellar winds. Additional mass
losses triggered by binarity would thus not induce more than a perturbation in
the
evolution
of these stars. In contrast, lower initial mass stars not suffering the LBV
winds
would be drastically affected by Roche Lobe Overflow mass transfer, which acts
as a
strong wind not operating if the corresponding stars were isolated.
 
The extent in the reduction of the \chem{26}{Al} that can be ejected in the
ISM by
$M > 40$ M$_\odot$ W--R members of binary systems cannot be assertained at this
point.
It has clearly to depend on the fraction of the mass lost by the W--R star
which
is
accreted by its companion, and thus withdrawn from the ISM. In this
scenario, however, one may wonder about the fate of the accreting companion,
and
about its
net production or destruction of \chem{26}{Al} resulting from its evolution
(see Langer et al. 1998).
Clearly, the effect of binarity on the net \chem{26}{Al} outcome by W--R stars
largely remains to be studied.

\section{THE METEORITIC CONNECTION}

The chemical composition of W--R
stellar winds may be relevant to the understanding of some isotopic
anomalies observed in meteorites (Arnould et al. 1997ab). In particular it
appears that a series of radionuclides have decayed in situ in some of them
(see
the review by Podosek \& Nichols 1997). This fact may be interpreted as
resulting
from the injection into the proto-solar nebula of these radioisotopes by one
or a few nucleosynthetic source(s). It is interesting
to mention that such an observation puts some constraints on $\Delta^*$, the
time elapsed
between the last astrophysical event(s) able to affect the composition of the
solar nebula and the solidification of some of its material. This time must be
shorter than the time required for the decay of the radionuclide if at least
part of it has to be trapped in live form in  meteorites. 

A detailed study of the production by non-rotating W--R stars of
short-lived radionuclides of astrophysical and cosmochemical interest has been
conducted by 
Arnould et al. (1997b). In short, their main results may be summarized as
follows:
  
%\begin{list}{(\arabic{obj})}{\usecounter{obj}\setlength{\leftmargin}{0pt}
%              \setlength{\labelwidth}{0pt}\setlength{\itemindent}{5pt}
%              \setlength{\listparindent}{\parindent}
%              \setlength{\parsep}{\parskip}
%              }
%\vskip-2truemm
%\item

\noindent (1) The neutrons released by \reac{22}{Ne}{\alpha}{n}{25}{Mg} during
the
He-burning phase of the considered stars are responsible for a s-type
process leading to the production of a variety of $A > 30$
radionuclides. In the absence of any chemical fractionation between the
relevant 
elements, it is demonstrated that
\chem{36}{Cl}, \chem{41}{Ca} and \chem{107}{Pd} can be produced by this
s-process in
a variety of W--R stars of the WC subtype with different initial masses and
compositions
at a {\em relative} level compatible with the meteoritic observations. For a
$60$ \ms $\,$ star with solar metallicity, Fig.~4 shows that this
agreement can be obtained for a time $\Delta^\ast \approx 2\times10^5$ y, where
$\Delta^\ast$ designates the time elapsed between the last
astrophysical event(s) able to affect the composition of the solar nebula and
the solidification of some of its material. More
details concerning other model stars are given by Arnould et al. (1997b);

%\item 
\noindent (2) To the above list of radionuclides, one of course has to add
\chem{26}{Al} (see above). The canonical
value $(\chim{26}{Al}/\chim{27}{Al})_0 = 5\times10^{-5}$ (MacPherson et al.
1995; the subscript $0$ refers to the start of the solidification sequence in
the solar system), while
not reached in the 60 \ms $\,$ star displayed in Fig.~4, can be obtained from
the
winds of $M \geq 60$ \ms $\,$ stars with $Z > Z_\odot$ under the same type of
assumptions as the ones adopted to construct Fig.~4. Let us also note that the
W--R
models can account for the correlation  between \chem{26}{Al} and \chem{41}{Ca}
observed in some meteorites  (Sahijpal et al. 1998);

%\item 
\noindent (3) Too little \chem{60}{Fe} is synthesized;
 
%\item 
\noindent (4) An amount of \chem{205}{Pb} that exceeds largely the experimental
upper
limit set by Huey \& Kohman (1972), but which is quite compatible with the
value
reported by
Chen \& Wasserburg (1987), is obtained not only for the model star displayed in
Fig.~4, but
also
for the other cases considered by
Arnould et al. (1997b). 

%\item
\noindent (5) More or less large amounts of \chem{93}{Zr}, \chem{97}{Tc},
\chem{99}{Tc} and \chem{135}{Cs} can also be produced in several cases, but
these predictions cannot be tested at this time  due to the lack of reliable
observations.
%\end{list}

It has to be remarked that the above conclusions are derived without taking
into
account the possible contribution from the material ejected by the eventual
SNIb/c
explosion of the considered W--R stars.
This SN might add its share of radionuclides that are not produced
abundantly enough prior to the explosion. This concerns in particular
\chem{53}{Mn}, \chem{60}{Fe} or \chem{146}{Sm}. One has also to acknowledge
that the above
conclusions
sweep
completely under the rug the possible role of rotation and binarity in the W--R
yields.  

>From the results reported above, one can try estimating if indeed there is
any chance for the contamination of the protosolar nebula with isotopically
anomalous W--R wind material at an {\em absolute} level compatible with the
observations. In the framework of Fig.~4, this translates into the possibility
of obtaining reasonable dilution factors $d(\Delta^\ast)$. A qualitative
discussion of this  highly complex question based on a quite simplistic
scenario is presented by Arnould et al. (1997b). In brief, it is concluded that
astrophysically plausible situations may be found in which one or several W--R
stars with masses and metallicities in a broad range of values could indeed
account for some now extinct radionuclides that have been injected live into
the forming solar system (either in the form of gas or grains). Of course, a
more definitive conclusion would have to await the results of a more detailed
model that takes into account the high complexity of the W--R circumstellar
shells, and of their interaction with their surroundings, demonstrated by
observation and suggested by numerical simulations. Concomitantly, the
possible role of W--R stars, either isolated or in OB associations, as triggers
of the formation of some stars, and especially of low-mass stars, should be
scrutinized.           

\beginrefer

\refer Arnould M, Prantzos N: 1999, {\it New Astronomy} {\bf 4}, 283

\refer Arnould M, Takahashi K: 1999, {\it Rep. Prog. Phys.} {\bf 62}, 395

\refer Arnould M, Meynet G, Paulus G : 1997a, in {\it Astrophysical
Implications of the Laboratory Study of Presolar Materials}, Eds. TJ
Bernatowicz \& EK Zinner, AIP {\bf 402}, p. 179

\refer Arnould M, Paulus G, Meynet G: 1997b, {\it Astron. Astrophys.} {\bf
321}, 452
 
\refer Azzopardi M, Lequeux J, Maeder A: 1988, {\it Astron. 
      Astrophys.} {\bf 189}, 34

\refer Braun H, Langer N: 1995, {\it Astron. Astrophys.} {\bf 297}, 483

\refer Chen JH, Tilton GR: 1976, {\it Geochimica et Cosmochimica Acta},
{\bf 40}, 635

%\refer Clayton D: 1982, in {\it Essays in Nuclear Astrophysics}, Eds. C.
%Barnes et al, Cambridge University Press, p. 401

\refer Conti PS, Garmany CD, de Loore C, Vanbeveren D: 1983, {\it Ap. J.}  
      {\bf 274}, 302

\refer Conti PS: 1984. In {\sl Observational Tests of the Stellar
      Evolution Theory},
      IAU Symp. {\bf 105}, Eds. A Maeder, A Renzini, Dordrecht: Reidel, pp. 233

\refer Crowther PA, Hillier DJ, Smith LJ: 1995,
      {\sl Astron. Astrophys.} {\bf 293}, 403

\refer De Marco O, Schmutz W: 1999, {\sl A\&A} {\bf 345}, 163

\refer Diehl R, Timmes FX: 1998, {\it Pub. Astron. Soc. Pacific} {\bf 110}, 637

\refer Hamann WR, Koesterke L: 1998, {\sl Astron. Astrophys.} {\bf 333}, 251

\refer Huey JM, Kohman TP: 1972, {\sl Earth Planet Sci. Letters} {\bf 16}, 401

\refer Kn\"odlseder J, Dixon D, Bennett K, et al: 1999, {\sl A\&A} 
{\bf 345}, 813

\refer Lamers HJGLM, Maeder A, Schmutz W, Cassinelli JP: 1991, 
      {\sl   Ap. J.} {\bf 368}, 538

\refer Langer N, Braun H, Fliegner J: 1995,
      {\sl Astrophys. and Space Science} {\bf 224}, 275

\refer Langer N, Braun H, Wellstein S: 1998, in Nuclear Astrophysics 9,
Eds. W. Hillebrandt \& E. M\"uller, MPA-report P10, p. 18

\refer MacPherson GJ, Davis AM, Zinner EK: 1995, {\sl Meteoritics} {\bf 30},
365

\refer Maeder A: 1981, {\sl A\&A} {\bf 102}, 401

\refer Maeder A: 1983, {\sl A\&A} {\bf 120}, 113

\refer Maeder A, Conti PS: 1994, {\sl ARAA} {\bf 32}, 227

\refer Maeder A, Lequeux J, Azzopardi M 1980: {\sl Astron. Astrophys.} {\bf
90}, L17

\refer Maeder A, Meynet G: 1994 {\bf A\&A} {\bf 287}, 803

\refer MacPherson GJ, Davis AM, Zinner EK: 1995, {\sl Meteoritics}
{\bf 30}, 365

%\refer Mahoney WA, Ling JC, Wheaton WA, Jacobson AS: 1984, 
%{\sl ApJ} {\bf 286}, 578

\refer Massey P, Johnson O: 1998, {\sl Ap. J.} {\bf 505}, 793

\refer Meynet G, Arnould M, Prantzos N, Paulus G: 1997, {\sl A\&A}
{\bf 320}, 460.

\refer Meynet G, Maeder A, Schaller G, Schaerer D, Charbonnel C: 1994,
{\sl A\&AS} {\bf 103}, 97

%\refer Mochizuki YS, Kumagai S, Tanihata I: 1999, in {\sl Origin of Matter and
%Evolution of Galaxies 97}, Eds. S. Kubono et al., World Scientific, p. 327

\refer Murty SVS, Goswami JN, Shukolyukov YA: 1997, {\sl Ap. J.} {\bf 475}, L65

%\refer Nagataki S: 1999, preprint (astro-ph/9907109)

%\refer Oberlack U, Bennett K, Bloemen H, et al: 1996, {\sl A\&AS} 
%{\bf 120}, 311C

\refer Podosek FA, Nichols RHJr:1997, in {\sl Astrophysical Implications
  of the Laboratory Study of Presolar Materials}, Eds. T.J. Bernatowicz \& E.K
Zinner, AIP {\bf 402}, p. 617

\refer Prantzos N, Diehl R: 1996 {\sl Phys. Rep.} {\bf 267}, 1

\refer Sahijpal S et al: 1998, {\sl Nature} {\bf 391}, 559

\refer Smith LF: 1973, In {\sl W--R and High-Temperature Stars}, IAU Symp. {\bf
49}, Eds.
      MKV Bappu, J Sahade, Reidel:Dordrecht, pp. 15

\refer Smith LJ: 1988, {\sl Ap. J.} {\bf 327}, 128

\refer Srinivasan G, Ulyanov AA, Goswani JN: 1994, {\sl Ap. J.} {\bf 431}, L67

\refer Takahashi K: 1998, in {\sl Tours Symposium on Nuclear Physics III},
Eds. M. Arnould et al., AIP Conf. Proceedings {\bf 425}, p. 616

\refer van der Hucht KA: 1992, {\sl Astron. Astrophys. Rev.} {\bf 4}, 123

\refer Willis AJ: 1991, in {\sl Wolf-Rayet Stars and Interrelations with
Other Massive Stars in Galaxies}, Proc. IAU Symp. {\bf 143}, K.A. van der Hucht
\& B. Hidayat (eds.), Kluwer, Dordrecht, p. 265

\refer Willis AJ: 1999. In {\it Wolf--Rayet Phenomena in Massive Stars and
      Starburst Galaxies}, IAU Symp. {\bf 193}, ASP Conf. Ser. Eds. KA 
      van der Hucht et al. pp. 1

\refer Willis AJ, Dessart L, Crowther PA, Morris PW, Maeder A,
      Conti PS, van der Hucht
      KA: 1997, {\sl MNRAS} {\bf 290}, 371

\refer Willis AJ, Dessart L, Crowther PA, Morris PW, van der Hucht KA:
      1998, {\sl Astrophys. and Space Sc.} {\bf 255}, 167

\refer Zahn, J-P: 1992, {\sl A\&A} {\bf 265}, 115

\endrefer           
\end{document}